\documentclass[12pt]{article}

\usepackage{scicite}
\usepackage{times}
\usepackage{graphicx}
\usepackage{amsmath}
\usepackage{amssymb}
\usepackage{amsfonts}
\usepackage{mathrsfs}

\newenvironment{sciabstract}{%
\begin{quote} \bf}
{\end{quote}}

\title{Quantitative assessment of the role of undocumented infection in the 2019 novel coronavirus (COVID-19) pandemic}

\author
{Yong-Shang Long,$^{1}$ Zheng-Meng Zhai,$^{1}$ Li-Lei Han,$^{1}$ Jie Kang,$^{2}$\\
Yi-Lin Li,$^{2}$ Zhao-Hua Lin,$^{1}$ Lang Zeng,$^{1}$ Da-Yu Wu,$^{1}$\\
Chang-Qing Hao,$^{1}$ Ming Tang,$^{1,3\ast}$ Zonghua Liu,$^{1}$ Ying-Cheng Lai$^{4\ast}$\\
\\ 
\normalsize{$^{1}$ State Key Laboratory of Precision Spectroscopy}\\
\normalsize{and School of Physics and Electronic Science,}\\ 
\normalsize{East China Normal University, Shanghai 200241, China}\\
\normalsize{$^{2}$ School of Communication and Electronic Engineering,}\\
\normalsize{East China Normal University, Shanghai 200241, China}\\
\normalsize{$^{3}$ Shanghai Key Laboratory of Multidimensional Information Processing,}\\
\normalsize{East China Normal University, Shanghai 200241, China}\\
\normalsize{$^{4}$ School of Electrical, Computer and Energy Engineering,}\\
\normalsize{Arizona State University, Tempe, AZ 85287, USA}\\
\\
\normalsize{$^\ast$ To whom correspondence should be addressed;}\\
\normalsize{E-mail: tangminghan007@gmail.com,Ying-Cheng.Lai@asu.edu}
}

\date{}

\begin{document}
\baselineskip24pt
\maketitle

\begin{sciabstract}

An urgent problem in controlling COVID-19 spreading is to understand the role of undocumented infection. We develop a five-state model for COVID-19, taking into account the unique features of the novel coronavirus, with key parameters determined by the government reports and mathematical optimization. Tests using data from China, South Korea, Italy, and Iran indicate that the model is capable of generating accurate prediction of the daily accumulated number of confirmed cases and is entirely suitable for real-time prediction. The drastically disparate testing and diagnostic standards/policies among different countries lead to large variations in the estimated parameter values such as the duration of the outbreak, but such uncertainties have little effect on the occurrence time of the inflection point as predicted by the model, indicating its reliability and robustness. Model prediction for Italy suggests that insufficient government action leading to a large fraction of undocumented infection plays an important role in the abnormally high mortality in that country. With the data currently available from United Kingdom, our model predicts catastrophic epidemic scenarios in the country if the government did not impose strict travel and social distancing restrictions. A key finding is that, if the percentage of undocumented infection exceeds a threshold, a non-negligible hidden population can exist even after the the epidemic has been deemed over, implying the likelihood of future outbreaks should the currently imposed strict government actions be relaxed. This could make COVID-19 evolving into a long-term epidemic or a community disease a real possibility, suggesting the necessity to conduct universal testing and monitoring to identify the hidden individuals.
	
\end{sciabstract}

\section*{Introduction}

At the end of December 2019, an unexplained, new type of coronavirus emerged
in the city of Wuhan in Hubei province of China. On January 10, 2020, the
health officials of Wuhan city confirmed, through nucleic acid detection, 41
cases. Coincided with ChunYun, the annual period of mass migration in China for
the Spring Festival holidays, the virus began to spread from Wuhan city to
other regions of China. On January 23, 2020, the Health Department of the
Central Chinese Government locked down Wuhan, a city of 11 millions, and a
number of other cities and regions in Hubei province, and at the same time
implemented rigorous, nationwide travel restrictions, in an attempt to prevent
large scale spreading of the disease~\cite{NHC:2020}. In spite of the strict
measures, suspected and confirmed cases began to emerge in almost every
province and major city of China, with a rapid increase in a short time
period. On February 11, 2020, the International Committee on Taxonomy of
Viruses designated the virus as severe acute respiratory syndrome coronavirus
2 (SARS-CoV-2) and on the same day, the World Health Organization (WHO)
officially named the disease as COVID-19. The unprecedented, rigorous measures
imposed by the Central Chinese Government have proven to be quite effective at
controlling and suppressing the spread, stabilizing its dynamics. At the time
of writing, there were only a few newly confirmed cases in China. While
spreading in China has apparently diminished, SARS-CoV-2 has begun to
emerge in counties and regions outside China, with large scale spreading
in South Korea, Italy, Iran, and now in the United States. With the likelihood
that COVID-19 would become a catastrophe in public health, on February 28
WHO has raised the global risk level of the new corona pneumonia epidemic
from ``high'' to ``very high.'' On March 11, WHO declared COVID-19 a global
pandemic.

For COVID-19, a number of data analysis studies have provided estimates of the
basic parameters underlying the spreading dynamics. In particular, the very
first 425 confirmed cases were analyzed~\cite{Lietal:2020} with the findings
that the average incubation period was 5.2 days, the doubling time was about
7.4 days and the basic reproduction number $R_0$ was about 2.2 extracted from 
the exponential growth behavior, and the infection due to personal
contacts had already occurred. An analysis~\cite{Novel:2020} of the 72314
confirmed cases in China up to February 11, 2020 found 889 asymptomatic cases
(about $1.2\%$) and 1023 deaths ($2.3\%$). From the growth curve,
it was extrapolated that the epidemic in China would peak about January 23-26.
An analysis of a data base of large number of samples
revealed~\cite{Guanetal:2020} the median incubation period of about four days,
quartile of five days (two-seven days), and death rate about $1.4\%$.
According to the joint China-WHO Joint Investigation Report on New Crown
Pneumonia, the median time interval between symptom appearance and confirmation
in China was 12 days (8-18 days) in January and reduced to three days
(1-7 days) at the beginning of February, but in Wuhan the respective numbers
are 15 days (10-21 days) and five days (3-9 days). Initial usable data
indicated that, the median time between mild symptoms and cure was about
two weeks, while that between severe symptoms and cure was between three
and six weeks, and it took about one week to go from initial symptoms to
severe illness such as hypoxia. Among the cases who did not survive, the
time from initial symptoms to death was between two and eight weeks. There
are similarities between SARS-CoV-2 and SARS-Cov (Severe Acute Respiratory
Syndrome associated Coronavirus) or MERS-Cov (Middle East Respiratory
Syndrome associated Coronavirus), but there are also
differences~\cite{Novel:2020}. While the propagation scenarios among
the three viruses are similar, the rapid spreading in China and evidence of
infection through human-to-human contacts suggested that SARS-CoV-2 is more
infectious than SARS-Cov or MERS-Cov, and the strong infectability during the
relatively long incubation period is particularly worrisome~\cite{WHHG:2020}.
These features of SARS-CoV-2, in spite of the associated relatively small
death rate, can cause more damage to the society, posing a significant
challenge to control, mitigation, and prevention.

For virus spreading, developing an effective mathematical model for making 
reliable predictions is of paramount importance to quantitatively assessing the
epidemic trend as well as to control and prevention. In dealing with recent
viruses of worldwide impact such as SARS-Cov~\cite{ZMB:2004},
MERS-Cov~\cite{CBSMV:2014}, Ebola~\cite{Ebola:2014} and
ZiKa~\cite{TBCFMR:2016} viruses, the modeling approach played an indispensable
role. For COVID-19, an SEIR (Susceptible-Exposed-Infectious-Recovered) model
was developed~\cite{WLL:2020} based on the OAG (Official Aviation Guide) data,
the inter-city population movement data from Tencent, and parameters values
with the incubation and infectious periods taken from SARS-CoV. Utilizing
cases outside of China, which were originated from Wuhan, to inversely deduce
the epidemic spreading process in major cities in China, the
authors~\cite{WLL:2020} estimated the $R_0$ value to be 2.68 and the doubling
time of 6.4 days. Further, it was predicted that by January 25, 2020, the
number of infected cases in Wuhan would be 75,815. This model is quite
effective, but two main deficiencies stood: the relevant parameter values were
from the previous epidemic of SARS-CoV and, the infectability during the
incubation period, perhaps the most distinct feature of SARS-CoV-2, was not
taken into account. The SEIR model was also used with parameter values from
People's Daily (the official Central Chinese Government newspaper) to estimate
the $R_0$ value based on the initial exponential growth~\cite{Zhouetal:2020},
with the result $R_0 \approx 2.8\sim 3.3$. Another study~\cite{Xuetal:2020}
based on the population movement data from Tencent and Baidu carried out a
statistical analysis of the geographic distribution of the population exiting
Wuhan, giving an assessment of the impact of this type of population on the
epidemic. An improved SEIR model~\cite{Yangetal:2020} taking into account the
infectability during the incubation period and incorporating machine learning
trained based on the data from 2003 SARS-CoV, predicted that the epidemic in
China would peak in the second half of February and stabilize at the end of
April. A remarkable result~\cite{Yangetal:2020} was that, should the Central
Government's rigorous quarantine and control measures be delayed for five
days, the size of the epidemic in China would be tripled and, if the lockdown
measures of Wuhan were relaxed, a second peak would emerge and last through
the second half of April. An assumption in this model was that the incubation
time distribution is Markovian, i.e., the occurrence time of the transition
from incubation to being infected follows the Poisson distribution with
exponentially distributed interevent time. However, for COVID-19, the assumed
Markovian spreading process is too idealized, as there were substantial
empirical data~\cite{Guanetal:2020} indicating that the incubation period of
COVID-19 has a strong non-Markovian characteristic, i.e., there is
a time delay between incubation and appearance of symptoms. A delayed
spreading model incorporating the non-Markovian characteristics was then
developed~\cite{Yueetal:2020}, taking into account the city lockdown and
using an inverse approach for parameter estimation, which predicted the
infection rate for mainland China of about 0.23 and isolation rate of about
0.42. A difficulty is that the current isolation measures in China were
implemented in response to the development of the epidemic, which are
time-dependent and may even be temporal. It has been
predicted~\cite{Guanetal:2020} that, had the implementation
of the government control measures been delayed for five days, the epidemic
scale in mainland China would have been three times larger. The effects of
travel restrictions in China on global COVID-19 spreading have also been
studied~\cite{Chinazzieaba9757}.

While the recent studies~\cite{Lietal:2020,Novel:2020,Guanetal:2020,WHHG:2020,
WLL:2020,Zhouetal:2020,Xuetal:2020,Yangetal:2020,Yueetal:2020} provided useful
insights into the COVID-19 pandemic, three significant characteristics were
not taken into account: (1) the existence of a non-zero fraction of hidden 
and undocumented individuals who carry the virus but are asymptomatic during 
incubation, mildly symptomatic, or even never symptomatic, (2) non-Markovian 
features associated with state transitions during the epidemic, and (3) the 
effects of time-dependent activities associated with population movements and 
contacts. The models constructed so far without taking into these key features 
generated parameter estimates that deviate from those of the real spreading 
dynamics, with incorrect predictions about the epidemic dynamics and trend. 
While the government provides the numbers of the newly confirmed and 
accumulated cases on a daily basis, it is impossible to know the accurate 
number of the hidden individuals who are infectious but have not been 
isolated or quarantined. These individuals are precisely the single most 
important factor for possible future outbreaks. Indeed, a very recent 
study~\cite{LPCSZYS:2020} revealed that substantial undocumented infection 
tends to facilitate the rapid spreading of COVID-19 as has been witnessed in 
many countries. There was speculation of the possibility of COVID-19 evolving 
into a community disease and a long term epidemic but there has been no 
supporting evidence so far. If, after the current epidemic is over, the 
fraction of virus carrying individuals in incubation approaches zero, then
the likelihood for a future outbreak would be very low. However, the
alternative devastating scenario could arise: if the fraction is nonzero and
persists after the current pandemic, the danger of a new outbreak and COVID-19
becoming a community disease would be real, and this would have indescribable
consequences to the whole world. The main goal of our work is to provide
quantitative evidence to either support or reject this scenario.

In this paper, we develop a five-state, non-Markovian spreading model for
COVID-19 incorporating the time delays associated with various state
transitions. The individuals who are asymptomatic are viewed as an important
reason for the currently ongoing large scale spreading. In addition, the
non-Markovian characteristics associated with state transitions and the
time-dependent variations in the human activities under the rigorous
governmental actions of isolation, quarantine, and travel restrictions
are fully accounted for. Simulations indicate that our model is capable of
accurately predicting the epidemic trend in China, South Korea, Italy, and
Iran. In fact, the large variations in the estimated number of the undocumented
individuals in different countries due to the disparity in government actions
have no effect on the predicted occurrence time of the inflection point. Our
model provides an explanation for the abnormally high mortality in Italy.
Based on the currently available data, our model predicts catastrophic 
epidemic scenarios in United Kingdom without strict travel and social 
distancing restrictions. As the number of confirmed cases approaches zero, 
the seemingly quiescent state may justify relaxation of the currently 
rigorously enforced quarantine policies (e.g., in China). However, such 
relaxation can lead to an increase in the fraction of hidden and undocumented 
individuals. Our model predicts that, if this fraction exceeds a certain 
value, the epidemic duration can increase by as long as two months, rendering 
likely a future outbreak. Another prediction is that the decay of the hidden 
population can be slower than that of the infected one and can maintain at a 
small but non-zero value for an extended period of time, and this has 
potentially devastating consequences: the activities of these individuals in 
combination with the unusually strong virulence of SARS-CoV-2 imply that they 
are a ``time bomb'' for a large scale outbreak in the future. The prolonged 
existence of the hidden/undocumented individuals makes COVID-19 evolving into 
a long term epidemic and a community disease a real possibility. Our findings 
provide the base for articulating and implementing further government actions, 
e.g., universal testing and monitoring to uncover and identify the individuals 
in the hidden state, to prevent future outbreaks.

\section*{Model and parameter estimation}

\subsection*{Five-state model}

\begin{figure} [ht!]
\centering
\includegraphics[width=0.9\linewidth]{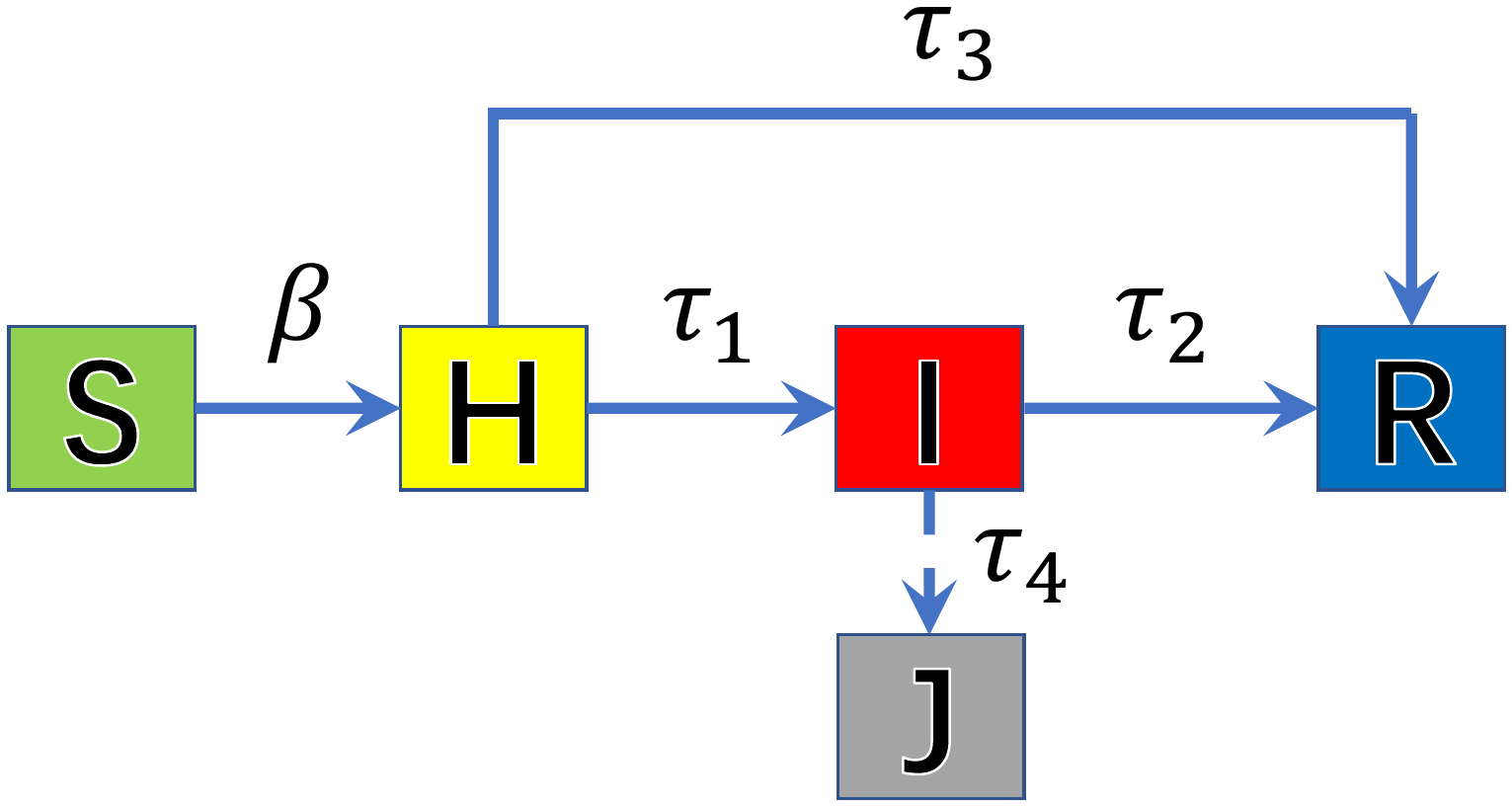}
\caption{ {\em Schematic illustration of COVID-19 spreading dynamics in
the five-state model}: S - susceptible, H - hidden, I - infected,
J - infected and confirmed, and R - recovered (cured or died). The parameter
$\beta$ is the infection rate, and $\tau_1$, $\tau_2$, $\tau_3$, and $\tau_4$
represent the four relevant time delays between different state pairs due to
the strong non-Markovian nature of the spreading dynamics.}
\label{fig:model}
\end{figure}

At each time step, an individual can be in one of the five states: susceptible
(S), hidden (H), infected (I), confirmed and isolated (J), and recovered (R).
An individual in S is healthy but can be infected. An individual in H carries
the virus but is asymptomatic or mildly symptomatic, and he/she can infect 
other individuals. There
are two types of individuals in the H state: those who are quarantined and
those who are not. Individuals in the I state have been infected and exhibit
symptoms. A fraction of the infected population have been confirmed and
hospitalized, to which we designate a new state, J. Finally, individuals in
the R state are either recovered or died of the disease. Because of the strong
government actions, the infected individuals are fully aware of their state
and are either quarantined or self-isolated; these individuals thus would not
spread the virus, which is especially true for those in the J state who are
in hospitals. Likewise, individuals in the R state are not infectious. Thus,
in our model, only the individuals in the H state are able to infect others
to spread the virus. Note that, one could define the H state as one that
contains the asymptomatic individuals and those who are symptomatic but have
not been identified. In this case, the duration of the H state consists of
the medical incubation period (from the time of infection to the time of
symptoms) and the time taken to isolation, while the I state contains
symptomatic and isolated individuals with a time between isolation and
confirmation. The distributions of the H and I time can be obtained from
empirical data.

The spreading dynamics are illustrated schematically in Fig.~\ref{fig:model}.
The susceptible individuals are infected at rate $\beta$ by those in the H
state who are not isolated. The S individuals who have been infected switch
to state H. Among the individuals in the H state, a fraction $\eta$ can
recover spontaneously through their own immune system or die, a process that
requires $\tau_3$ days. (Parameter $\eta$ is thus the fraction of undocumented
cases.) The remaining $(1-\eta)$ fraction of the H state individuals
exhibit symptoms and switch to the I state after an average incubation period
of $\tau_1$ days, due to the non-Markovian nature of symptom appearances.
Individuals in the I state are subject to medical treatment and will either
recover or die after $\tau_2$ days - the average time for the transition
from I to R states. A unique feature of our model is the introduction of the
J state: on average the I individuals will need $\tau_4$ days to be confirmed.
Mathematically, these processes can be described by the following set of
time-delayed dynamical equations:
\begin{eqnarray} \label{eq:s}
\frac{dS(t)}{dt} & = & -\Phi(t), \\ \nonumber
\frac{dH(t)}{dt} & = & \Phi(t)-(1-\eta)\int_{0}^{t}f_1(\tau)\Phi(t-\tau)d\tau-{\eta}\int_{0}^{t}f_3(\tau)\Phi(t-\tau){d}\tau \\ \label{eq:h}
	& & -(1-\eta)f_1(t)H(0)-{\eta}f_3(t)H(0), \\ \nonumber 
\frac{dI(t)}{dt} & = & (1-\eta)\int_{0}^{t}f_1(\tau)\Phi(t-\tau)d\tau
-(1-\eta)\int_{0}^{t}f_2(\tau')d\tau'\int^{t-\tau'}_{0}f_1(\tau)\Phi(t-\tau'-\tau){d}\tau \\ \label{eq:i}
	& + & (1-\eta)f_1(t)H(0) - (1-\eta)\int^{t}_{0}f_2(\tau)f_1(t-\tau)H(0)d\tau - f_2(t)I(0), \\ \nonumber
\frac{dR(t)}{dt} & = & \eta\int_{0}^{t}f_3(\tau)\Phi(t-\tau){d}\tau
	+ (1-\eta)\int_{0}^{t}f_2(\tau')d\tau'\int^{t-\tau'}_{0}f_1(\tau)\Phi(t-\tau' - \tau){d}\tau \\ \label{eq:r}
	& + & \eta f_3(t)H(0) + (1-\eta)\int^{t}_{0}f_2(\tau)f_1(t-\tau)H(0)d\tau + f_2(t)I(0), \\ \nonumber
\frac{dJ(t)}{dt} & = & (1-\eta)\int_{0}^{t}f_4(\tau')d{\tau'}\int_0^{t-\tau'}f_1(\tau)\Phi(t-\tau'-\tau){d}\tau \\ \label{eq:j}
	& & +(1-\eta)\int_{0}^{t}f_4(\tau)f_1(t-\tau)H(0)d\tau,
\end{eqnarray}
where $\Phi(t)=\beta S(t)H_l(t)/N$ is the growth rate of the H-state
population and $H_l(t)=l(t)H(t)$ is the number of hidden individuals in the H
state who have not been quarantined with $l(t)$ quantifying the travel
activity level of the population which decays exponentially with time [See
Supplementary Information (SI)]. The quantities $f_1(\tau)$, $f_2(\tau)$,
$f_3(\tau)$ and $f_4(\tau)$ are probability distributions of the incubation
period ($\tau_1$), of the treatment time ($\tau_2$), of the self-healing time
period for the hidden individuals ($\tau_3$), and of the time period for an
infected individual to be diagnosed ($\tau_4$), respectively. The distribution
functions are assumed to be normal~\cite{du2020serial}:
$f_i(\tau)=(\sqrt{2\pi}\sigma_i)^{-1}\exp{[-(\tau-\mu_i)^2/(2\sigma_i^2)]}$
for $i=1,2,3,4$, with $\mu_i$ and $\sigma_i$ denoting the mean and standard
deviation of the distribution, respectively.

In Eq.~\eqref{eq:h}, the first term on the right-hand side is the inflow rate
of the number of newly emerged individuals in the H state, the second (third)
term represents the outflow rate from the H state of the individuals who
have been hidden for time $\tau$ to the I (R) state, and the fourth (fifth)
term is the outflow rate of the individuals who are initially in the H state
and switch to the I (R) state. For Eq.~\eqref{eq:i}, the first term on the
right side is the inflow rate of individuals from the H state to the I state,
the second term denotes the outflow rate of the new individuals in the H state
switching to the I state after time $\tau$ and then to the R state after time
$\tau'$, the third term represents the inflow rate of the individuals
initially in the H state into the I state, the fourth term is the outflow rate
of the individuals who are initially in the H-state, switch to the I state
after time $t-\tau$, and then change to the R state after time $\tau$, and the
fifth term represents the outflow rate of the individuals who are initially
in the I state and are recovered after time $t$. The various terms in
Eqs.~\eqref{eq:r} and \eqref{eq:j} can be interpreted in a similar way.

\subsection*{Inference of model parameter values}

Based on the joint China-WHO Joint Investigation Report on New Coronavirus
Pneumonia and Ref.~\cite{Guanetal:2020}, we set the average incubation
period to $\mu_1 = 5 \ \mbox{days}$, the average time from infection
to recovery ($\tau_2$) to be $\mu_2=18 \ \mbox{days}$. For the value of
the fraction $\eta$ of undocumented infection, due to the variations among
different countries in terms of the criteria for testing and confirming
infected cases and because of the difficulty to measure this parameter,
we make a reasonable assumption about its value. Likewise, because of the
lack of empirical data to assess the average time $\mu_3$ of spontaneous
recovery, we exploit the fact that these individuals are more immune to
SARS-CoV-2 than the average population and accordingly set $\mu_3=14$. 
From empirical data, we have $\mu_4 = 7$, but it has little effect on the 
dynamics.

In addition to $\eta$, which other model parameters are key to the current
spreading dynamics of COVID-19, subject to government actions? The following
features of COVID-19: long hidden time, strong virulence, mild symptoms for
many individuals, and difficulty in detecting the SARS-CoV-2 virus, strongly
suggest that the population in the H state are mainly responsible for spreading
the disease. In addition, the direct impact of the government actions is an
exponential reduction in the human movement activities. The parameters $H(0)$,
$\beta$, and $\lambda$ are thus key, which can be estimated based on the
relatively accurate daily number $J(t)$ of the confirmed cases as reported by
the governments. In particular, the values of the three parameters can be
inferred from the following least squares optimization method:
\begin{equation} \label{eq:LM}
	\min_{\beta,H(0),\lambda} ||J(\beta,H(0),\lambda) - J_{data}||_2.
\end{equation}
We use the Levenberg-Marquardt (LM)
method~\cite{Levenbert:1944,Marquardt:1963,KNS:book} to solve
(\ref{eq:LM}) to obtain the optimal values $\beta^*$, $H^*(0)$,
and $\lambda^*$.

\section*{Results}

\subsection*{Determining key model parameters for different countries}

\begin{table} [ht!]
\centering
\caption{Optimal estimates of key parameters of COVID-19 spreading dynamics for China, South Korea, Italy, and Iran}
\label{tab:parameters}
\begin{tabular}{|c|c|c|c|c|c|}
\hline
 & Fit for 25 days (China) & Global optimal (China) & Korea & Italy & Iran\\
\hline
$\beta^{*}$&0.38&0.37&0.27&0.29&0.31\\
\hline
$H^{*}(0)$&460&525&91&35&800\\
\hline
$\lambda^{*}$&0.16&0.15&0.18&0.18&0.18\\
\hline
\end{tabular}
\end{table}

Table~\ref{tab:parameters} lists the values of the three key model parameters:
$\beta$, $H(0)$, and $\lambda$, for five different countries, where 
$\eta = 0.012$ for China and $\eta = 0.1$ for South Korea, Italy, and Iran.
Another important parameter is the fraction $\eta$ of undocumented infection
that reflects on the effectiveness of the government actions. The ways by
which these parameters are obtained differ for different countries. In China,
the value of $\eta$ has been officially announced: $\eta = 0.012$. For South
Korea, Italy, and Iran, no official report of the value of $\eta$ is
available, so we simulate the model for both small and large values of $\eta$.
For a given value of $\eta$, the optimal values of $\beta$, $H(0)$, and
$\lambda$ are obtained through the LM optimization method (\ref{eq:LM}) in
combination with integration of Eqs.~(\ref{eq:s}-\ref{eq:j}).

To explain the meaning of parameter $\lambda$, we take China as an example.
The lockdown of Wuhan and other cities in the Hubei province occurred
on January 23. There were 12 days of active travel: between January 11 and
22. The travel activity function $l(t)$ is thus given by
\begin{equation} \label{eq:activity}
l(t) = \left\{
\begin{aligned}
& 1, \ \  t < t_0, \\
& e^{-\lambda (t - t_0)}, \ \ t_0 \le t \leq t_0 + T, \\
& e^{-\lambda T}, \ \ t \ge t_0 + T
\end{aligned}
\right.
\end{equation}
where $t_0 = 12$. Since the spreading dynamics in China was triggered by the
human outflow from Wuhan, with the implementation of the strict governmental
control measures nationwide, virus spreading in cities other than Wuhan can be
neglected. We thus have the total number of individuals participating in the
spreading dynamics as $N = 14$ millions. Because of the lockdown of Wuhan,
the human activity function $l(t)$ decreases exponentially but it cannot be
zero. We thus set $l(t)$ to be $e^{-\lambda T}$ (a small constant) for
$T = 7$ days after the lockdown.

\begin{figure} [ht!]
\centering
\includegraphics[width=\linewidth]{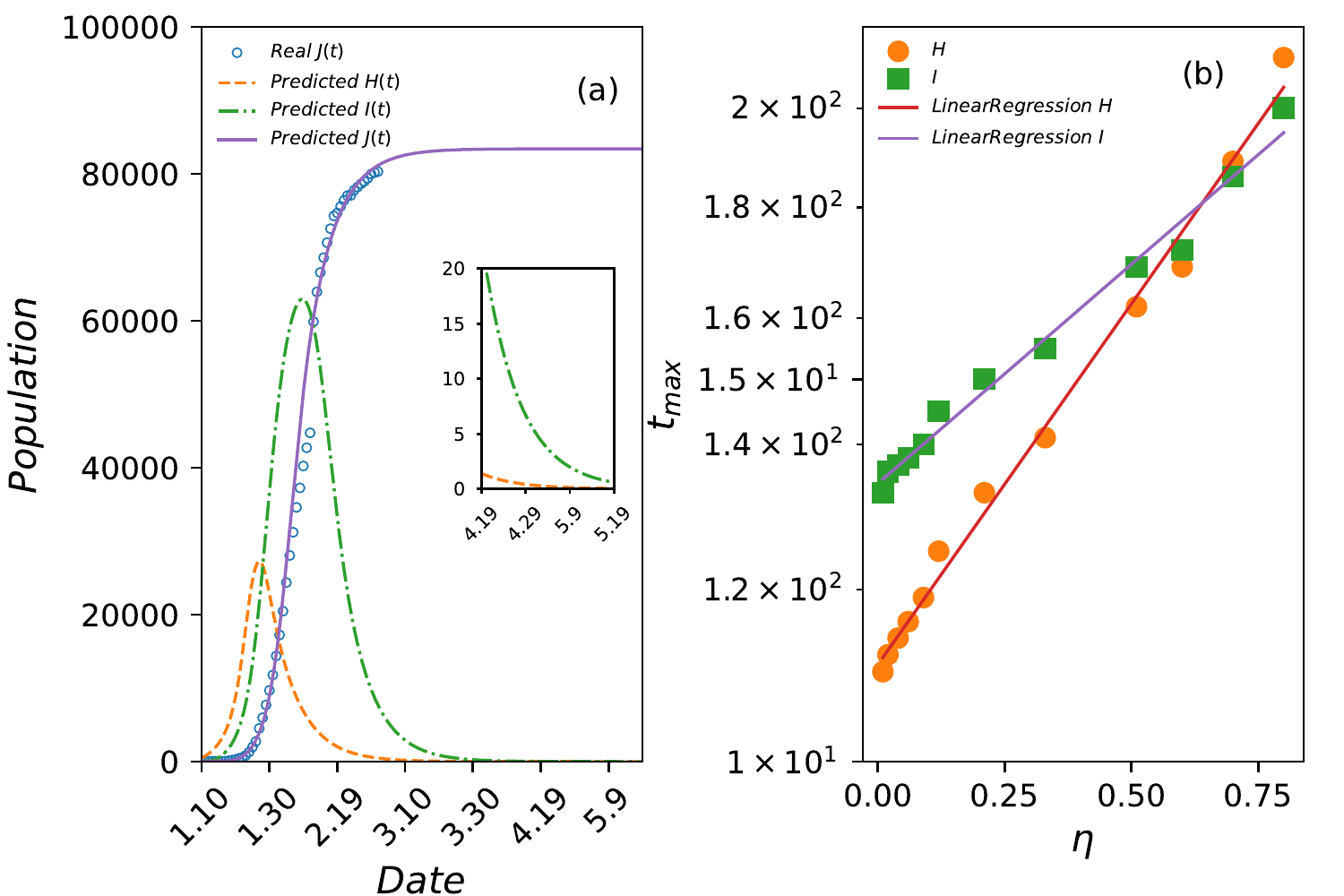}
\caption{ {\em Prediction of COVID-19 epidemic trend in China}.
(a) The purple curve is the predicted daily number of confirmed cases in China
between January 10 and May 9, which agrees well with the available actual
data points (open blue circles). The green dot-dashed and orange dashed curves
represent the predicted numbers of infected and hidden individuals,
respectively. The value of $\eta$ is taken to be 0.012, the official value
from the Chinese Epidemic Disease Control Center. The numbers of H-state
and I-state individuals are predicted to vanish around May 15, as shown
in the inset, which agree with the current epidemic trend in China. (b) The
number of days required for H-state (orange filled circles) and I-state
(green filled squares)  individuals to approach zero versus $\eta$. If the
value of $\eta$ is 0.75, the number of days required for the epidemic to end
would be about 200. At $\eta \approx 0.6$, a crossover between the two
lines occurs: after that the H state population lasts longer than that of
the I state, setting the stage for a possible second outbreak.}
\label{fig:China}
\end{figure}

\subsection*{Prediction for China}

We first demonstrate that our five-state model has the power to accurately
predict the epidemic trend of COVID-19 spreading in China in a detailed and
quantitative way, and present a striking finding. From the data of COVID-19
spreading in China, we set January 10, 2020 to be the reference time ($t = 0$)
with the initial conditions $I(0) = 41$, $J(0) = 41$, and $R(0) = 0$. With the
estimated parameter values and the initial conditions, we numerically solve
the set of delay differential equations Eqs.~(\ref{eq:s}-\ref{eq:j}).
Figure~\ref{fig:China}(a) shows, in a 100-day period (from January 10 to April
29), the predicted daily accumulated number of confirmed cases (the purple
curve) and the available actual data to date (open blue circles). The
agreement is remarkable, attesting to the predictive power of our model.
Figure~\ref{fig:China}(a) also shows the predicted daily numbers of the
I-state (green dot-dashed curve) and H-state (orange dashed curve) individuals.
It can be seen that the epidemic peaked on February 7-8, indicating the
occurrence of the inflection point. However, since the average time interval
between infection and confirmation is $\tau_4$, the observed occurrence of
the inflection point would be delayed by $\tau_4$, i.e., February 14-15,
which was indeed the time reported by the Chinese government. The epidemic
cycle is predicted to end towards the end of March when the number of infected
individuals approaches zero, which agrees with the currently observed epidemic
trend in China.

The above prediction of the epidemic trend in China in terms of $H(t)$ and
$I(t)$ is based on the choice $\eta = 0.012$, i.e., among all the H-state
individuals, only $1.2\%$ are undetected and spontaneously recovered or died 
eventually, which is reasonable, considering the extremely strict monitoring
and quarantine policies currently in place in China. If the government actions
were not that strict or it they were not rigorously enforced, the value of
$\eta$ could be much larger. In that case, how long will the epidemic last?
Our model was designed to answer such important questions. 
Figure~\ref{fig:China}(b) shows the number of days of the epidemic cycle, 
defined as the time required for $I(t)$ to approach zero, versus $\eta$, 
where an approximately exponential relationship is observed. It can be seen 
that, if the value of $\eta$ is $0.75$, the epidemic duration could be 200 
days. This means that, had the Chinese government not taken the unprecedentedly
dramatic steps to contain the COVID-19 spreading, the epidemic cycle could 
easily last into the summer. Figure~\ref{fig:China}(b) also shows the model 
predicted time required for the hidden, virus-carrying population to vanish 
versus $\eta$, which is also approximately exponential. The striking 
phenomenon is that the exponential rate is larger than that with the infected 
population. The two lines intersect at $\eta \approx 0.6$, indicating that, 
if $\eta$ exceeds $60\%$, the time for the H-state population to vanish can 
be longer than that required for the I-state population to diminish. Should 
the government actions be withdrawn when it is determined that $I(t)$ is 
already practically zero, the continuous existence of the remaining H-state 
population could be the source for a future outbreak! This group of 
virus-carrying individuals in the hidden state can diffuse into any place 
in the country or in the world, and they represent a ``time bomb'' for a new 
epidemic if the control and preventive measures of the government are 
completely withdrawn. The possible existence of this group strongly suggests 
continuous government actions even beyond the end of the current epidemic. 
Perhaps, COVID-19 is so unique and different from other viruses, rendering 
necessary universal testing and monitoring nationwide.

\begin{figure} [ht!]
\centering
\includegraphics[width=\linewidth]{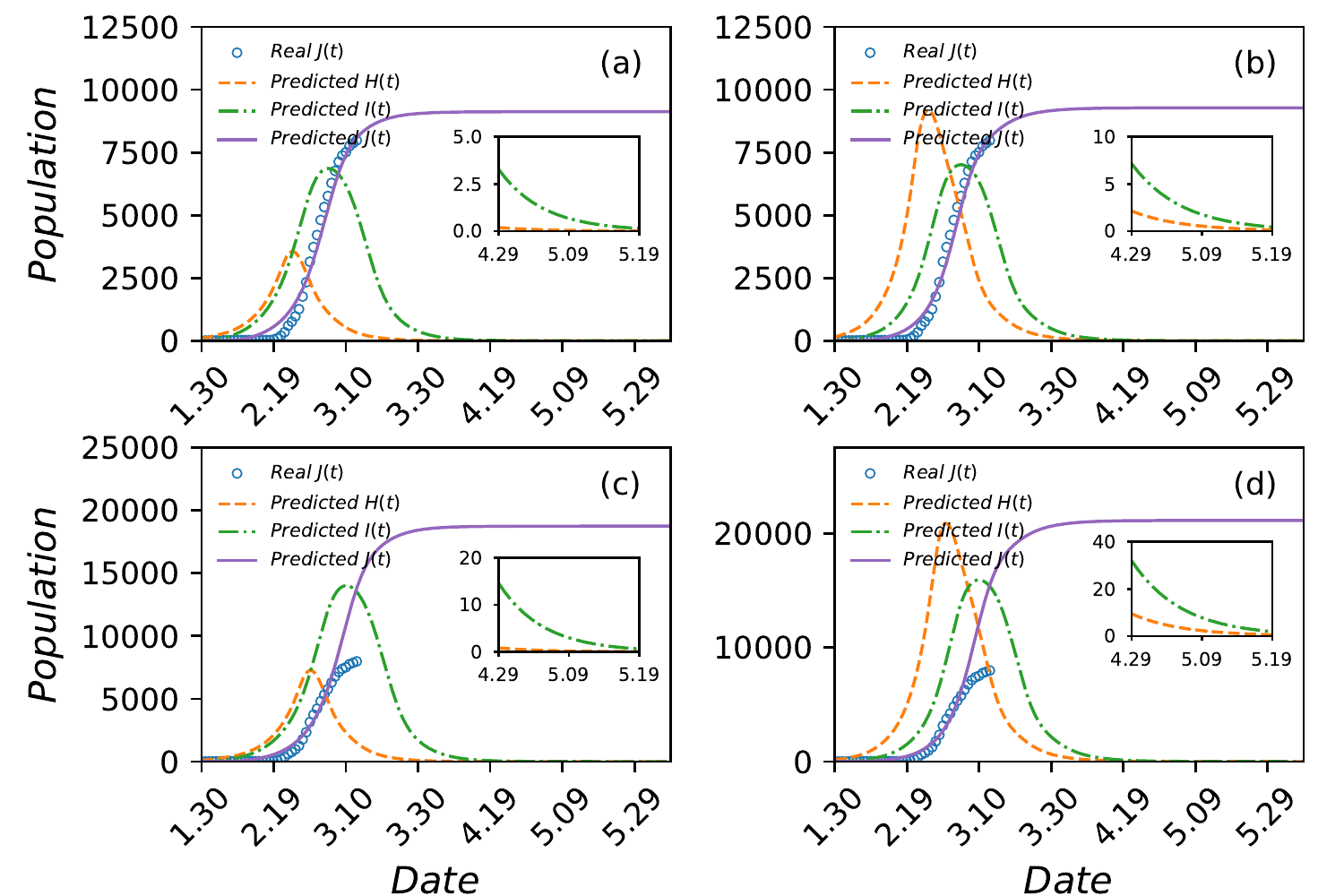}
\caption{ {\em Model prediction of COVID-19 epidemic trend in South Korea}.
(a) For $\eta = 0.1$ [$(\lambda^*,H^*(0),\beta^*) = (0.18,91,0.27)$],
the predicted behaviors of $H(t)$, $I(t)$ and $J(t)$ and the real data of
$J(t)$. (b) Same plots as in (a) but for $\eta = 0.5$ [
$(\lambda^*,H^*(0),\beta^*) = (0.26,155,0.23)$]. In both cases, the model
predictions agree with the current data. (c,d) The predicted epidemic
behaviors in the hypothetical scenario that the lockdown of the epicenter
were delayed for five days, where other parameter values are the same as
those in (a,b), respectively. The delay would cause the final size of
the infected population to be doubled and lead to a somewhat prolonged
epidemic duration.}
\label{fig:Korea}
\end{figure}

\subsection*{Prediction for South Korea}

For South Korea, we use the reported data~\cite{WHO:2020} of the number of 
confirmed cases between January 30 and March 13. The epidemic emerged mainly 
in the cities of Daegu and Gyeongsangbuk-do, so we set $N = 5$ millions (the 
approximately combined population of the two cities). The two cities, being the
epicenter, were locked down on February 21, so we have $t_0 = 21$. The value
of $\eta$ is not available, so we test two different values: 0.1 and 0.5,
with the values of the other three model parameters estimated to be
$(\lambda^*,H^*(0),\beta^*) = (0.18,91,0.27)$ and
$(\lambda^*,H^*(0),\beta^*) = (0.26,155,0.23)$, respectively. The prediction
results for the two sets of parameter values are shown in
Figs.~\ref{fig:Korea}(a) and \ref{fig:Korea}(b), respectively. In spite of
the difference in the choice of the value of parameter $\eta$, in both cases
our model predicts the following: occurrence of the epidemic peak (inflection
point) on March 5$\sim$7, the final number of confirmed cases of about 9200,
and the epidemic duration of approximately three months. These predictions
agree with the current data in South Korea. For the two choices $\eta = 0.1$
and $\eta = 0.5$, there is a small difference in the predicted ending date of
the epidemic: about May 6 for the former and May 13 for the latter. We also
simulate the hypothetical scenario where the lockdown of the two cities were
delayed for five days, with the results for $\eta = 0.1$ and $\eta = 0.5$
shown in Figs.~\ref{fig:Korea}(c) and \ref{fig:Korea}(d), respectively.
Inevitably, the predicted curves of the accumulative number of confirmed
cases deviate from the data, but the final epidemic size would double and
the epidemic could last longer.

\subsection*{Prediction for Italy: Why is the mortality so high?}

\begin{figure} [ht!]
\centering
\includegraphics[width=\linewidth]{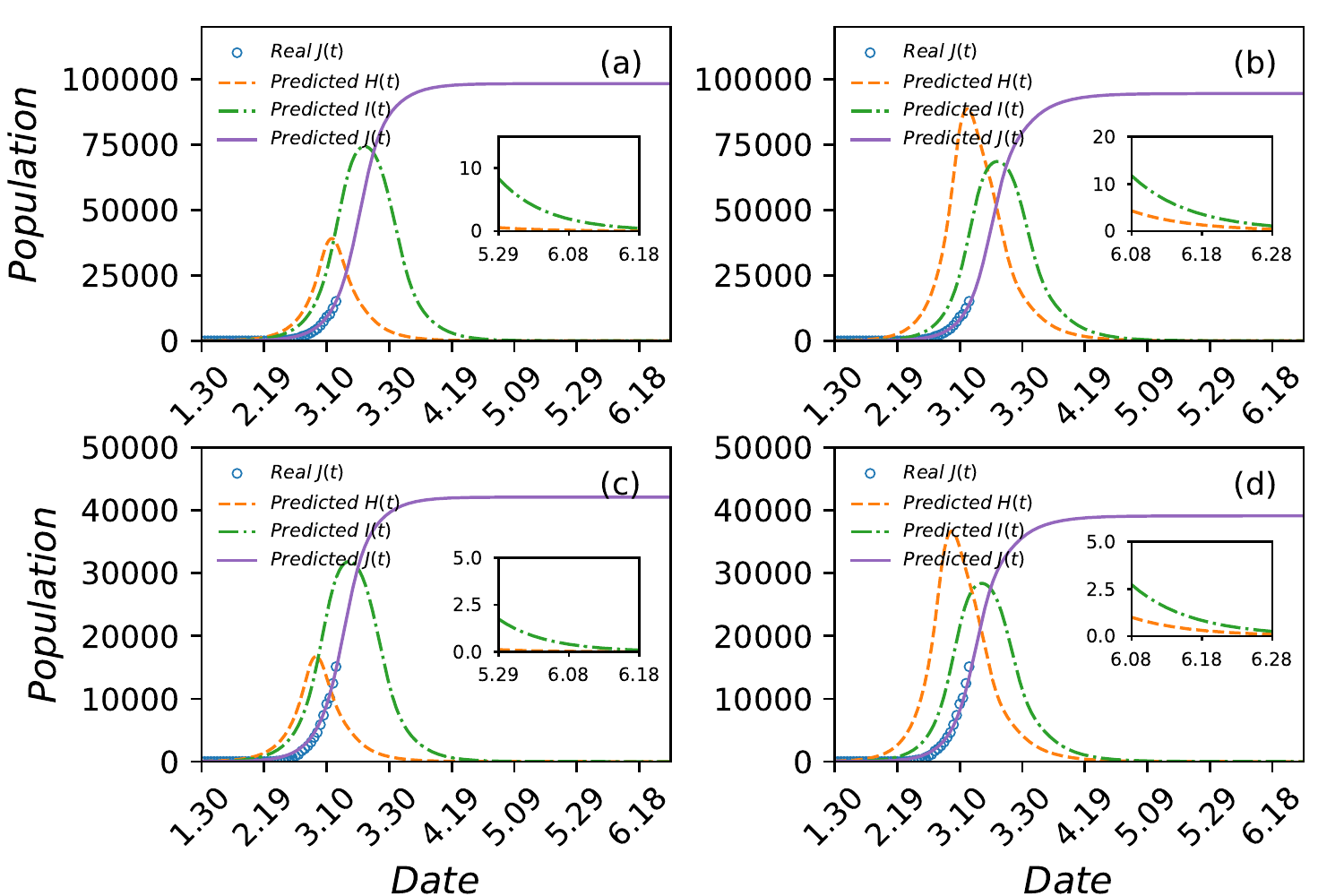}
\caption{ {\em Model prediction of COVID-19 epidemic trend in Italy assuming
the same lockdown effects as in China}. (a) For $\eta = 0.1$
[$(H^*(0),\beta^*) = (35,0.29)$], the predicted
behaviors of $H(t)$, $I(t)$ and $J(t)$ and the real data of $J(t)$.
The lack of sufficient data after the lockdown of the country prevents a
reliable estimate of the value of $\lambda$, so we assume a similar effect
as in China and use $\lambda = 0.18$ as in Fig.~\ref{fig:China}(b).
(b) For $\eta = 0.5$ [$(H^*(0),\beta^*) = (65,0.24)$], the predicted
results, where $\lambda = 0.24$ as in Fig.~\ref{fig:China}(b).
In both cases, the model predictions agree with the current data.
(c,d) The predicted epidemic behaviors under the hypothetical scenario that
the country were locked down five days earlier for the same parameters
in (a,b), respectively. The result of an earlier lockdown would reduce
the final epidemic size by approximately a factor of two.}
\label{fig:Italy_1}
\end{figure}

\begin{figure} [ht!]
\centering
\includegraphics[width=\linewidth]{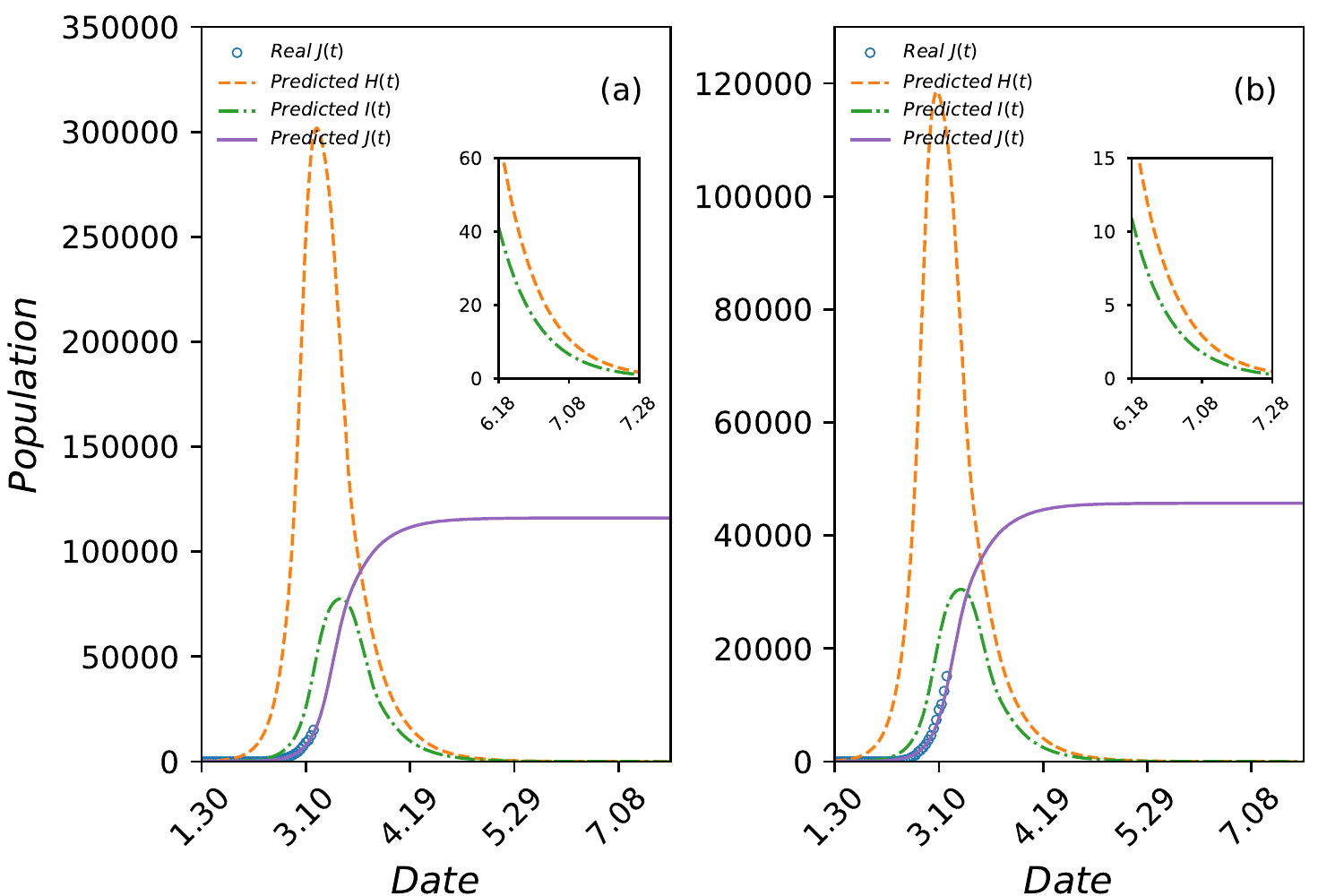}
\caption{ {\em Model prediction of COVID-19 epidemic trend in Italy
taking into account the actual effect of lockdown}.
The actual situation in Italy is that only $20\%$ of the infected individuals
are medically treated (versus nearly $100\%$ in China), leading to a larger
value of $\eta$. (a) For $\eta = 0.8$ [$(H^*(0),\beta^*) = (130,0.22)$]
and $\lambda = 0.24$, the predicted behaviors of $H(t)$, $I(t)$ and $J(t)$
and the real data of $J(t)$. For this more realistic parameter setting,
the predicted peak epidemic size and hidden population would be about 75,000
and 300,000, respectively. (b) The predicted results under the hypothetical
scenario of a five-day early lockdown, where the epidemic figures would be
reduced by more than a factor of two.}
\label{fig:Italy_2}
\end{figure}

For Italy, we use the reported number of confirmed cases between January 30
and March 13. The entire country was locked down on March 8, so we have
$t_0 = 38$. The total population of Italy is $N \approx 60$ millions.
Figures~\ref{fig:Italy_1}(a) and \ref{fig:Italy_1}(b) present the model
prediction results of $H(t)$, $I(t)$, and $J(t)$, together with the actual
up-to-date $J(t)$ data, for $\eta = 0.1$ and $\eta = 0.5$, respectively.
There is a good agreement between the predicted and true behaviors of $J(t)$.
Figures~\ref{fig:Italy_1}(c) and \ref{fig:Italy_1}(d) display the prediction
results corresponding to the parameter settings in Figs.~\ref{fig:Italy_1}(a)
and \ref{fig:Italy_1}(b), respectively, under the hypothetical scenario that
the country were locked down five days earlier. It can be seen that the
effect of an earlier lockdown would reduce the final epidemic size
approximately by a factor of two.

There is a key difference between the lockdown of Wuhan in China and that of
Italy. Especially, in China, confirmed individuals are treated at hospitals
but in Italy, $80\%$ of such individuals are isolated in their home
(undocumented) and only $20\%$ are treated, leading to a relatively large
value of $\eta$. The predicted results in Figs.~\ref{fig:Italy_1}(a) and
\ref{fig:Italy_1}(b), where the value of $\eta$ is relatively small, thus may
not reflect the actual epidemic behavior. To remedy this deficiency, we
carry out prediction for $\eta = 0.8$, with results shown in
Fig.~\ref{fig:Italy_2}(a). It can be seen that the predicted $J(t)$ curve
still agrees with the actual data, but the predicted epidemic size now stands
at about 75,000 and the predicted peak H-state population is an astounding
300,000. Simulation reveals that there will still be hundreds infected and
hidden individuals after June 8. Our model predicts that, for Italy, for the
more realistic case of $\eta=0.8$, the epidemic will end at the beginning of
August (around August 3), which is about 50 days later than that for the case
of $\eta=0.1$. A striking feature is that, toward the end of the epidemic,
there are more H-state than I-state individuals, posing a significant
challenge to control and prevention. The prediction that the H-state
population can last longer than the I-state population is particularly
worrisome, because it implies a higher likelihood of a future outbreak.
Figure~\ref{fig:Italy_2}(b) shows the result for the hypothetical scenario of
a five-day earlier lockdown, where both the peak I-state and H-state
populations are more than halved, and the numbers of the two remaining
populations would be less than 200 by May 19.

A puzzling phenomenon that has been widely noticed and discussed is the
markedly higher mortality in Italy than that in South Korea, both being
developed countries at the similar level. (As of March 18, the mortality
among those infected is $8.34\%$ in Italy but it is only $0.998\%$ in South
Korea.) Our model prediction for large values of $\eta$ provides a reasonable
explanation. In particular, while the predicted number of confirmed cases
does not depend strongly on the value of $\eta$, the case of $\eta=0.8$,
which more closely describes the current effects of government actions in
Italy, has devastating consequences. Especially, $\eta=0.8$ means that
about $80\%$ of the virus-carrying individuals eventually enter into the R
state after a time delay $\tau_3$, without going through any medical
treatment. Since the population in the R state consists of both recovered
individuals and deaths, under the assumption that the death rate with COVID-19
is invariant across different countries, the significantly large fraction
of people switching directly from the H state to the R state means
significantly more deaths.

\subsection*{Prediction for Iran}

\begin{figure} [ht!]
\centering
\includegraphics[width=\linewidth]{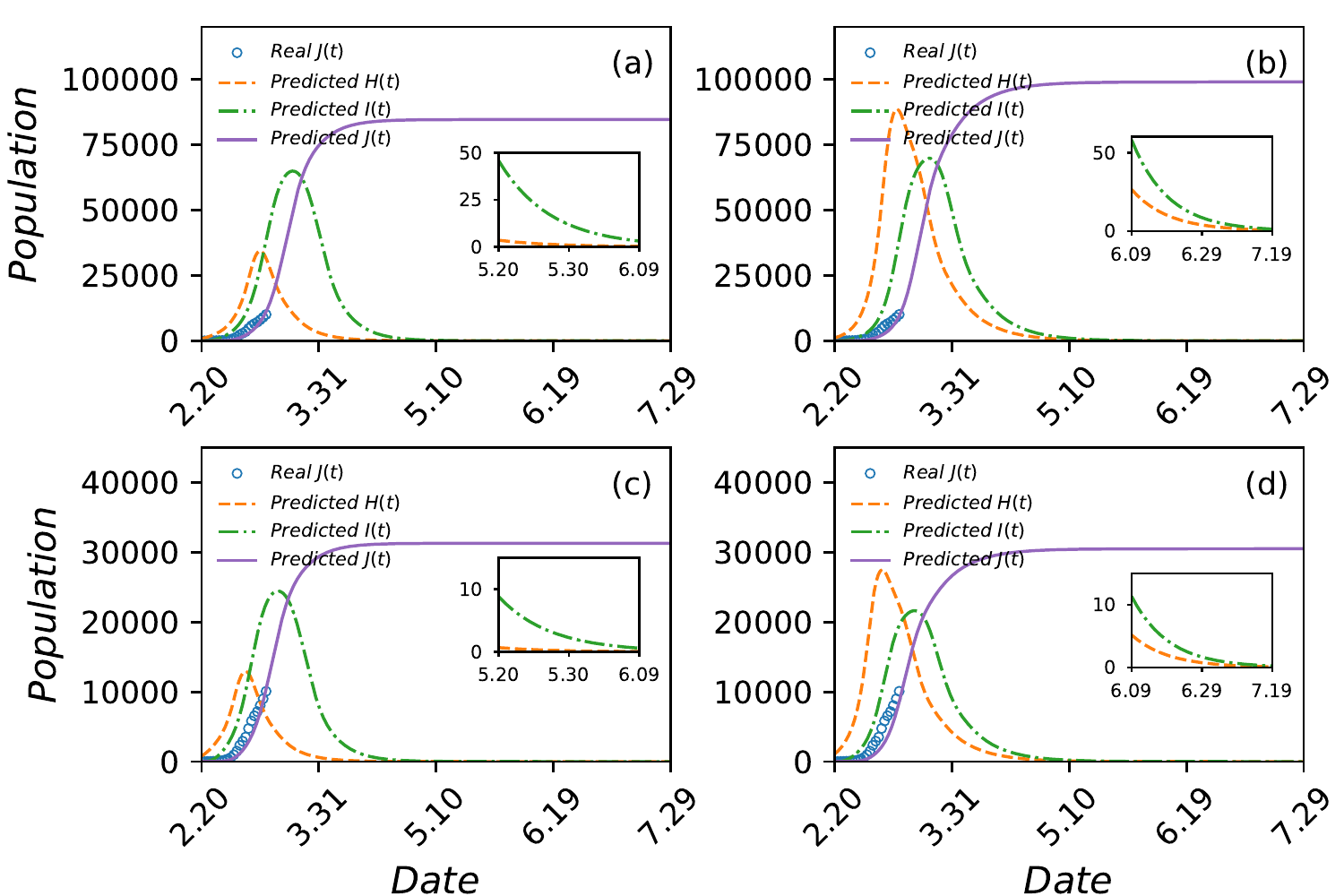}
\caption{ {\em Model prediction of COVID-19 epidemic trend in Iran}.
(a) For $\eta = 0.1$ [$(H^*(0),\beta^*) = (800,0.31)$], the predicted
behaviors of $H(t)$, $I(t)$ and $J(t)$ and the real data of $J(t)$. The value
of $\lambda$ is taken to be the same for China: $\lambda=0.18$ as in
Fig.~\ref{fig:China}(a), assuming similar lockdown effects between the two
countries. (b) For $\eta = 0.5$ [$(H^*(0),\beta^*) = (1100,0.29)$], the
predicted results, where $\lambda = 0.24$ as in Fig.~\ref{fig:China}(b).
In both cases, the model predictions agree with the current data.
(c,d) The predicted epidemic behaviors under the hypothetical scenario that
the country had been locked down five days earlier for the same parameters
in (a,b), respectively. The result of an earlier lockdown would significantly
reduce the epidemic size and shorten its duration.}
\label{fig:Iran}
\end{figure}

For Iran, lockdown occurs on March 7, so we have $t_0 = 16$. The total
population is $N \approx 88$ millions. Model predicted results are shown
in Figs.~\ref{fig:Iran}(a) and \ref{fig:Iran}(b) for $\eta = 0.1$ and
$\eta = 0.5$, respectively. In both cases, the infected population will
reach its peak around March 23. As for other countries, the value of $\eta$
does not affect the prediction of the inflection point. It can also be
seen that, for a larger value of $\eta$ [Fig.~\ref{fig:Iran}(b)], 
the number of confirmed individuals is greater than that for the case in
Fig.~\ref{fig:Iran}(a), and the duration is longer: for the former
the epidemic is predicted to end on about July 21 while it is June 17 for
the latter. Figures~\ref{fig:Iran}(c) and \ref{fig:Iran}(d) present the
respective model predicted results for $\eta = 0.1$ and $\eta = 0.5$ under
the hypothetical scenario of imposing lockdown five days earlier, where the
epidemic size could be significantly smaller with a much shorter duration.

\subsection*{Prediction of epidemic scenarios for United Kingdom}

\begin{figure} [ht!]
\centering
\includegraphics[width=\linewidth]{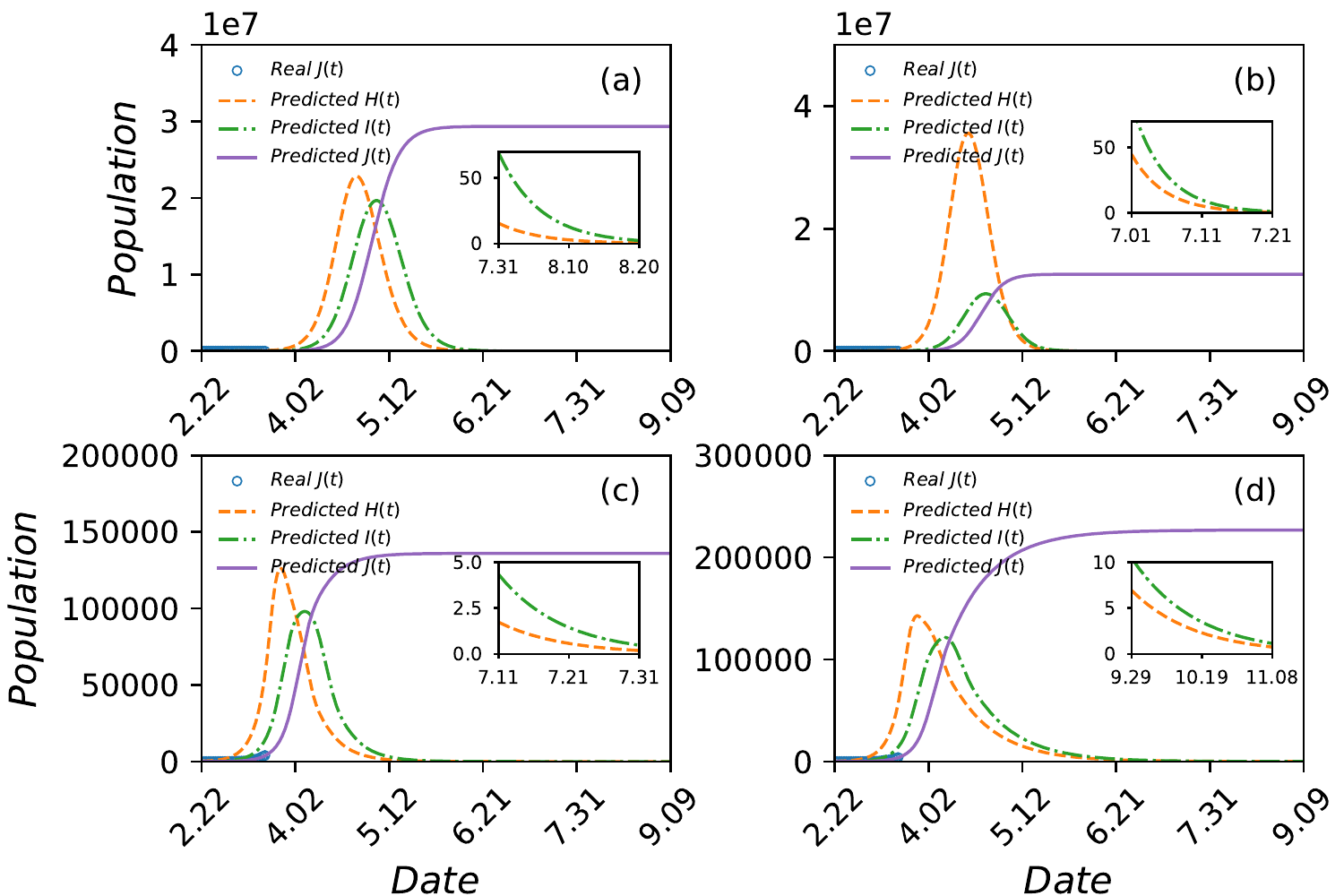}
\caption{ {\em Model prediction of COVID-19 epidemic scenarios in United
Kingdom}. Model prediction for four scenarios: (a) $\eta = 0.5$, $H(0) = 220$, 
and no government restrictions; (b) $\eta = 0.8$, $H(0) = 400$, and no
government restrictions; (c) $\eta = 0.5$, $H(0) = 220$, and with government 
restrictions as strict as those in China: $\lambda = 0.24$; (d) $\eta = 0.5$, 
$H(0) = 220$, and with less strict government restrictions: $\lambda = 0.18$. 
In all cases, the estimated infection rate is $\beta = 0.26$. Scenario (c) 
would result in the least damage, but scenario (d) better fits the current 
situation in United Kingdom.}
\label{fig:UK}
\end{figure}

For United Kingdom, sparse data became available on February 1 with a few
imported cases of infection, but the data after February 22 are more 
systematic. Model parameter optimization reveals that choosing February 1
or February 22 as the starting date of epidemic would lead to a different 
value of $H(0)$, but the value of $\beta$ is hardly affected: in both cases 
we have $\beta = 0.26$. The total population of the country is $N = 66$ 
millions. We simulate four scenarios. 
\begin{enumerate}
\item
Moderate fraction of undocumented cases ($\eta = 0.5$) and absence of  
government imposed travel and social distancing restrictions. The model 
predicted results are shown in Fig.~\ref{fig:UK}(a): $80\%$ of the population 
will be infected with approximately 30 million confirmed cases. The 
inflection point will occur around May 15 and the epidemic will be over
on about August 20.
\item
High fraction of undocumented cases ($\eta = 0.8$) and absence of 
government restrictions. The results are shown in Fig.~\ref{fig:UK}(b): 
$90\%$ of the population will be infected with approximately 10 million 
confirmed cases. The inflection point will occur around April 25 and the 
epidemic will be over on about July 20.
\item
Moderate fraction of undocumented cases ($\eta = 0.5$) and strict 
government imposed restrictions at a level similar to that of China: 
$\lambda = 0.24$ (the same value in China for $\eta = 0.5$). The results
are shown in Fig.~\ref{fig:UK}(c): approximately $1.3\times 10^5$ confirmed 
cases, occurrence of inflection on about April 3, and end of epidemic at 
the end of July or beginning of August.
\item
Moderate fraction of undocumented cases ($\eta = 0.5$) but with less 
strict government imposed restrictions: $\lambda = 0.18$. As shown in 
Fig.~\ref{fig:UK}(d), in this case, the eventual number of confirmed cases 
will be about $2.2\times 10^5$, the inflection will occur on about April 7, 
and the epidemic will stretch into the beginning of November before it is over. 
\end{enumerate}
Based on the available data at the time of writing, the last scenario 
appears to fit with the situation in United Kingdom.

\section*{Discussion} \label{sec:discussion}

We have developed a realistic, five-state epidemic spreading model with
time delays for COVID-19, taking into account virulence of individuals in
incubation, the non-Markovian characteristics associated with the various
state transitions, and the exponential decay of population activity level
under strict government actions. Our model requires four parameters to simulate
the spreading dynamics: the fraction $\eta$ of undocumented virus-carrying
individuals spontaneously recovered or died, the effective infection rate,
the initial size of the population in incubation, and the rate of reduction
in human activities due to government actions. These parameters can be
obtained from government reports, reasoned, or estimated through a mathematical
inverse optimization approach. The model is capable of real-time prediction,
and has been validated as its prediction of the number of daily accumulative
confirmed cases agrees remarkably well with the current data. The effects of
government actions, as measured by the parameter $\eta$, can vary greatly
among different countries. For example, in South Korea, almost all possible
individuals exposed to the virus went through nucleic-acid tests, while
such tests are not being conducted in countries such as Japan, the United
States, and United Kingdom, leading to abnormally low number of confirmed
cases. Countries such as Italy are perhaps somewhere in between, so a large
fraction of asymptomatic/undetected virus-carrying individuals without going
through any medical treatment exists. In this case, the epidemic size would
be significantly larger with a prolonged duration. This not only has provided
a natural explanation for the abnormally high mortality in Italy, but also
implied a devastating picture for the epidemic trend: even after the epidemic 
has been deemed over, there can still be a small population of asymptomatic
individuals in the society, making a future outbreak and COVID-19 evolving
into a long epidemic or a community disease a real possibility.

Our predicted scenarios for United Kingdom indicate that the epidemic would 
be catastrophic without government imposed travel and social distancing 
restrictions. The current restrictions would lead to about 220,000 confirmed 
cases and a prolonged epidemic duration into November. This prediction is 
consistent with that by the Imperial College COVID-19 Response 
Team~\cite{IC_Report_2020}. The best scenario leading to the least damage 
is when the government imposed restrictions as strict as those in China.  

COVID-19 is unusual because even individuals in incubation can be highly
infectious yet they themselves can be asymptomatic. The existence of even
a small fraction of such individuals with an arbitrarily long incubation time
in the population (H-state individuals), can be extremely worrisome, as they
carry the virus and are capable of spreading it to the general public yet
they appear healthy in every reasonable way and thus may never be identified
without dramatic government actions. Right from the beginning of the COVID-19
pandemic, there was a concern that it could evolve into a community disease
and will always be with us. However, this was merely a speculation without
any quantitative justification. Our work has provided theoretical and modeling
based evidence for the likelihood of COVID-19 becoming a community disease.
This has grave consequences and significant implications. For example, for
the current COVID-19 epidemic in China, our model predicts a few dozen such
H-state individuals. As the government measures are gradually withdrawn,
these individuals could diffuse into the general population, leading to the
next COVID-19 epidemic/pandemic. But the currently implemented and enforced
epidemic control policies cannot be maintained indefinitely for economical,
social and other reasons. What should the government do then? Before the
successful development of an effective and feasible vaccine, one possibility
is to conduct universal testing of the entire population to identify the
remaining H-state individuals. Having said that, we are not in a position
to provide a solution to this problem. Rather, our goal was to generate
scientific support for government policies on controlling and preventing
future large scale COVID-19 epidemic, which we believe has been achieved
in the current work.

Our model takes into account the effects of government imposed control
and preventive measures and has been demonstrated to have the predictive
power for the epidemic trend. The discovery of the sustained existence of
a group of individuals in the hidden state provides the base for future policy
making. Because of the vast difference among different countries and even
among different regions in the same country in terms of factors such as
medical facilities and the effectiveness of the government actions, it is
necessary to carry out a more detailed analysis of the probability
distributions of the time delays associated with the relevant state
transitions, and to investigate how the changes in the distributions affect
the inflection point, epidemic size and duration. In China, there
is now evidence that many cases of mutual spreading occur in closed
environments such as families, factories, and even prisons, which are not
affected by travel restriction. The effects of such spreading scenarios
on the epidemic need to be studied. Another issue is to investigate
COVID-19 spreading dynamics on larger spatial scales. In particular, our
model is especially suited for an isolated city such as Wuhan, where
the current description and understanding of the spreading dynamics
subject to rigorous implementation of government measures are reasonable.
However, as the range of epidemic increases, the effect of large scale
population movements on the spreading dynamics must be studied, possibly
through the approach of network modeling with subpopulation dynamics.
As the epidemic spreading begins to weaken or diminish, government
actions will inevitably be relaxed. How to effectively prevent
a second epidemic is an urgent problem. Also, is it possible to articulate
time-dependent control and preventive measures that are adaptive to the
development of the epidemic? And how to minimize the extent of control
and travel restriction without sacrificing the expectations is another
issue worth immediate attention.

\section*{Data Availability}

All relevant data are available from the authors upon request.

\section*{Code Availability}

All relevant computer codes are available from the authors upon request.

\section*{Acknowledgments}

This work was supported by the National Natural Science Foundation of China
(Grant Nos. 11975099, 11575041, 11675056 and 11835003), the Natural Science
Foundation of Shanghai (Grant No. 18ZR1412200), and the Science and Technology
Commission of Shanghai Municipality (Grant No. 14DZ2260800). YCL would like
to acknowledge support from the Vannevar Bush Faculty Fellowship program
sponsored by the Basic Research Office of the Assistant Secretary of Defense
for Research and Engineering and funded by the Office of Naval Research
through Grant No.~N00014-16-1-2828.

\section*{Author Contributions}

Y.-S.L., M.T. and Y.-C.L. designed research; Y.-S.L., Z.-M.Z., L.-L.H, J.K., and Z.-H.L. performed research; Y.-L.L., Z.-H.L., L.Z., D.-Y.W., and C.-Q.H. contributed analytic tools; Y.-S.L., Z.-M.Z., M.T., Z.L., and Y.-C.L. analyzed data; M.T., Z.L., and Y.-C.L. wrote the paper.

\section*{Competing Interests}

The authors declare no competing interests.

\section*{Correspondence}

To whom correspondence should be addressed. E-mail: tangminghan007@gmail.com; Ying-Cheng.Lai@asu.edu


\end{document}